\documentclass[sigconf]{acmart}
\usepackage{multirow}
\usepackage{float}
\AtBeginDocument{%
  \providecommand\BibTeX{{%
    \normalfont B\kern-0.5em{\scshape i\kern-0.25em b}\kern-0.8em\TeX}}}

\setcopyright{acmcopyright}
\copyrightyear{2020}
\acmYear{2020}
\acmDOI{10.1145/1122445.1122456}

\acmConference[UbiComp '20]{UbiComp '20}{September 12--16, 2020}{Cancun, Mexico}

\acmPrice{15.00}
\acmISBN{978-1-4503-XXXX-X/18/06}

\begin{document}

\title{Sensor Data and the City: Urban Visualisation and Aggregation of  Well-Being Data }


\author{Thomas Johnson}
\email{thomas.johnson@ntu.ac.uk}
\affiliation{%
  \institution{Nottingham Trent University}
  \city{Nottingham}
  \country{UK}
}

\author{Eiman Kanjo}
\email{eiman.kanjo@ntu.ac.uk}
\affiliation{%
  \institution{Nottingham Trent University}
  \city{Nottingham}
  \country{UK}
}

\author{Kieran Woodward}
\email{kieran.woodward@ntu.ac.uk}
\affiliation{%
  \institution{Nottingham Trent University}
  \city{Nottingham}
  \country{UK}
}

\renewcommand{\shortauthors}{Johnson, Kanjo and Woodward}

\begin{abstract}
The growth of mobile sensor technologies have made it possible for city councils to understand peoples' behaviour in urban spaces which could help to reduce stress around the city. We present a quantitative approach to convey a collective sense of urban places. The data was collected at a high level of granularity, navigating the space around a highly popular urban environment. We capture people’s behaviour by leveraging continuous multi-model sensor data from environmental and physiological sensors. The data is also tagged with self-report, location coordinates as well as the duration in different environments. The approach leverages an exploratory data visualisation along with geometrical and spatial data analysis algorithms, allowing spatial and temporal comparisons of data clusters in relation to people’s behaviour. Deriving and quantifying such meaning allows us to observe how mobile sensing unveils the emotional characteristics of places from such crowd-contributed content.
\end{abstract}


\maketitle
\section{Introduction}

Stress-related illnesses are drastically increasing around the world. Stress is often difficult to detect and is commonly associated with a stigma of embarrassment and humiliation preventing people from receiving help or treatment. However, the impact of stress is profound as it costs the UK economy £2.4 billion per year due to work absence \cite{Perkbox2018}. To help alleviate stress, many people restore their attention and seek relief through meditation or outdoor recreation. Nature and urban environments offer a restorative experience to improve an individuals’ well-being. However, some environments are busy, hectic and stressful resulting in a negative impact in well-being and not relieving stress or exhaustion. With recent technological advances, several studies have emerged which can be utilized to assess the effects of built environments on humans using physiological sensors. For example, Heart Rate (HR) monitors and Electrodermal Activity (EDA) sensors, have shown enhanced results following the exposure to restorative environments. 

When comparing walking in the countryside compared with urban environments the results illustrated how places are constructed through different senses and people’ s bodies impact whether the place is perceived as welcoming and pleasant or hostile and aggressive \cite{Thrift2003}. Positive impressions about places encourage people to visit, these places are most commonly quiet, restful and tranquil allowing people to reduce their stress levels by providing a palliative to the nonstop attentional demands of typical, city streets. It is important to use objective sensor data to study the relationship between different places and the mental states of people who visit these places.  Kaplan et al. \cite{Kaplan1995}, Kohlleppel et al. \cite{Kohlleppel2002} have studied the power of natural environments to provide a restful experience encouraging a quick and strong recovery from any stress encountered. In addition to natural settings, coffee shops, health clubs, video arcades and some retail shops were shown to encourage the restoration from stress and can promote positive emotions  \cite{Rosenbaum2009}. 

Many studies have been conducted to study the effects of environmental noise on mental health and human well-being. The research results proved that noise can impair productivity and cause serious health problems such as chronic stress and heart diseases \cite{Stansfeld2003}. Physiological signals contain useful patterns that aim to help identify individual's mental state. For example, using physiological sensors such as Electrocardiogram, Electromyogram, skin conductance and respiration has found to determine a driver's overall stress levels \cite{Healey2005}. Furthermore, recent studies have begun to use physiological sensors to identify an individual's mood and emotions \cite{Woodward2018b} \cite{Woodward2019} \cite{Woodward2020}. 

There are many applications emerging in the area of environmental monitoring and the impact the environment has on health and well-being \cite{Kanjo2018a} \cite{Donaire-Gonzalez2019}. For instance, a wearable, low power, air quality and environmental monitoring sensor sampling a range of air pollutants (CO, NO2 and O3). The sensors in the system using real-time information helped those suffering from health related problems such as asthma which in turn helped those individual's to avoid highly-polluted environments \cite{Zappi2012}. Furthermore, the application NoiseSpy is a low-cost sound measurement device that monitors environmental noise levels, allowing users to explore an environment while visualising noise levels in real time \cite{Kanjo2010}. 

The relationship between emotional stability and an urban environment is especially important and challenging in cities. This is because the environments are typically highly dynamic and densely populated. The ubiquitous nature of smartphones coupled with sensors and increased computational power has allowed them to be considered as serious competitors to dedicated sensor platforms. Mobile phones and physiological sensors present many opportunities to observe human emotion in urban environments. \cite{Alajmi2013}.

\begin{figure}[h]
\centering
  \includegraphics[width= 6cm]{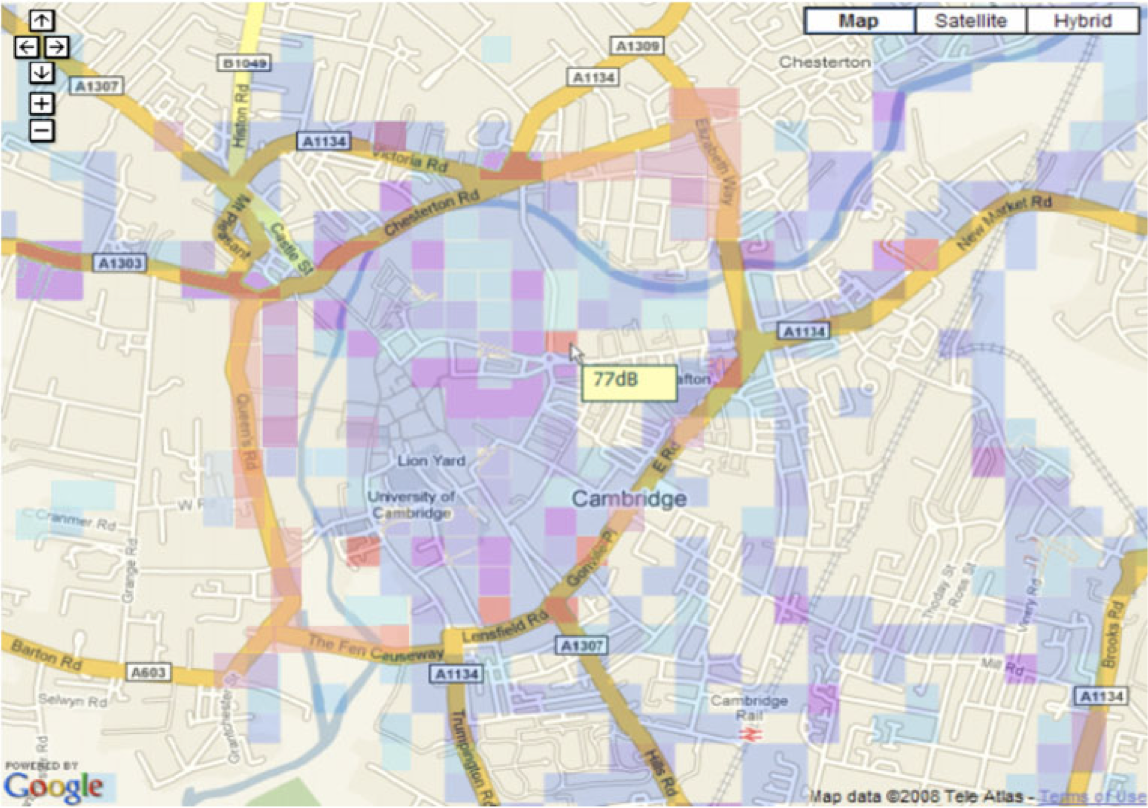}
  \caption{Visualising environmental noise data using grid overlay around Cambridge City.}
  \label{fig:mapgrid}
\end{figure}

Previous research on visualising sensor-based data spatially is focused primarily on mapping, heat-maps or grid overlay. The complexity of a spatial interpolation has a signiﬁcant impact on how much sensor data can be placed over a map. For example, Kriging can be used to visualise and map spatial data \cite{Krige1952}. Alternatively, heat maps can be utilised for a quicker visualisation. This has been used to track Electrodermal activity (EDA) in supermarkets and around shopping centres to understand how these environments have a negative impact on an individual's well-being  \cite{Alajmi2013} \cite{ElMawass2013}. Figure \ref{fig:mapgrid} presents a noise visualisation map whereby noise was recorded over a two week period. The darker colours on the grid view indicate a higher noise level in that particular location \cite{Kanjo2010}. 

There has been limited research focused on understanding how human emotion impacts when exploring an urban environment. A study \cite{Matei2001} has started to use this method by mapping human emotion around a place, by visualising on a map individual feelings (fear and comfort) in Los Angeles. However, although this is an early study in this area, the data was recorded prior and did not report in real-time which would have given a deeper understanding into the overall impact. The use of self-reporting to record well-being is becoming increasingly popular \cite{Woodward2020a} especially through mobile systems such as Mappiness \cite{MacKerron2012} and WiMO \cite{Mody2009} because of their ability to link the individual's emotion to a particular location.

We explore the use of physiological and environmental sensors in an urban city context to monitor how the environment impacts stress. The built environments considered in this work consist of both green spaces and urban built area allowing us to understand how the exposure to natural green spaces may promote greater attention restoration and stress recovery than visiting built environments. We propose a new method of spatial exploration and visualisation to help reduce stress within urban environments.

\section{Spatial Aggregation and Visualisation}
To visualise the dynamic trends, the urban area needs to be divided in smaller areas. Binning is a frequently used method to visualise geographical patterns which use heat maps of the data. Figure \ref{fig:mapgrid} illustrates the heat map of the physiological sensor data from one user data along within the city environment. The heat map shows that stressful hotspots are scattered along the path and not confined to one particular area.

\begin{figure}[h]
\centering
  \includegraphics[width= 7cm]{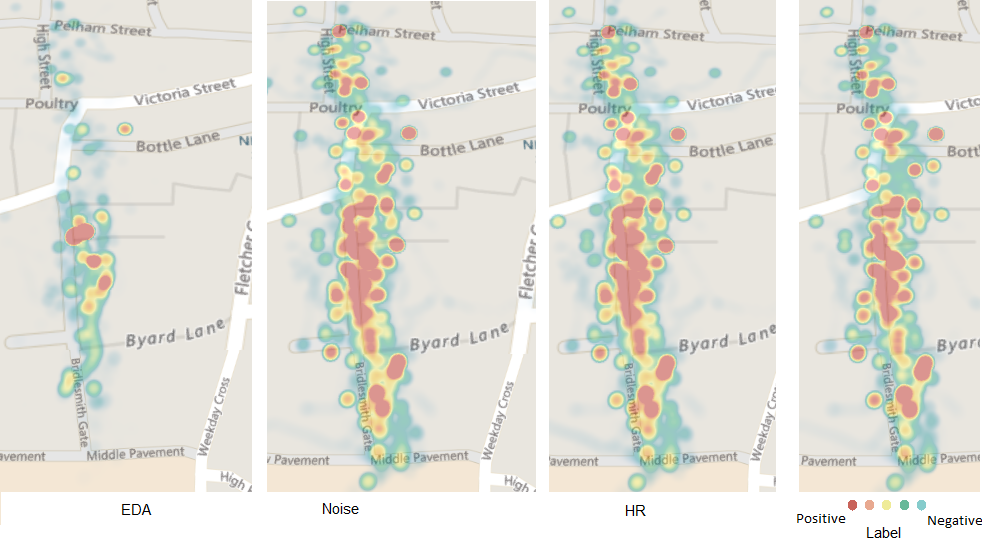}
  \caption{Map showing heat maps of various sensor data in  urban settings.}
  \label{fig:mapgrids}
\end{figure}

While Figure \ref{fig:mapgrids} shows the level of intensity of different urban environments based on GPS coordinates, the sensor data on these heat maps do not show the real distribution of sensor data. One option to solve this challenge is to divide the study area into grid cells \cite{Kanjo2010}; however, it is difficult to allocate a cell for each sensor reading, moreover, it is not possible to decide on the cell size, since the density of the sensor mobility traces can be of different density distributions.
 
The biggest challenge in metric layer visualisation lies in scalability. There can be large number of records to map in a single city. Drawing all these records at their accurate locations simultaneously will be impossible, due to the large amount of rendering cost and the required visual bandwidth.

To address these issues, we use a combinatorial computational geometry algorithm called “Voronoi”, which is a diagram partitioning a plane into regions based on distance to points in a specific subset of the plane \cite{Du1999}. Voronoi diagrams divide the space into a set of regions called Voronoi cells, including the space that is closest to the object (location). The size of these cells give an indication of the density of the area a certain object is in or the size of an object.

The cell structure also shows the Delaunay triangulation, which enables the calculation of an object's immediate set of neighbours. The definition of a Voronoi cell is given by the following equation, where x is a planar metric space; p is the set of generator points in the metric space; and d is the distance between all points in x and a specific generator point.

\begin{equation}
\text {Vori}=\{x \mid d(x, p i) \leq d(x, p j), j \neq i\}
\end{equation}

The creation of a Voronoi tessellations is a dynamic procedure that is repeated until all of the points are represented in adjacent polygons. If there are not a sufficient number of particles to satisfy Equation (1), then the Voronoi diagram is only partially filled and the data is redistributed. The Voronoi diagram is composed of a collection of tessellations defined as Vor, where:

\begin{equation}
V o r=\left\{V o r_{1}, V o r_{2} \ldots \operatorname{Vor}_{n}\right\}
\end{equation}

By giving each polygon a value that corresponds to the sensor value collected in a particular GPS coordinate, it is possible to divide the space into adjacent polygons with different sensor reading which are represented in colours. This type of spatial analysis, helps in understanding the degree to which a place is similar to other nearby places, the correlation among different spatial objects needs to be deﬁned. We have mapped users to different regions based on their mobility and form several user groups for further exploration. 

\section{Experimental setup}
We use the “EnvBodySens” dataset consisting of various sensor and self-report data from a smart phone application, and a Microsoft wristband 2. Collected data is logged and stamped with the time and date and GPS location. In the EnvBodySens application, an interface is implemented for continuous and quick labelling of user emotions while walking and collecting data. When the user launches the application, mobile interface appears with five iconic facial expressions ranging from very negative to very positive. Users were asked to constantly select one of the affective categories (in the form of buttons) as they walk around the city centre. We adopted the 5-step SAM Scale for Valence taken from \cite{Banzhaf2014} to simplify the continuous labelling process.

Forty participants took part in the study all female with an average age of 28. Participants were scheduled to take part in the study in order to collect data around Nottingham city centre. The street we selected for the experiment is a pedestrianised shopping street in the centre of Nottingham. The street hosts some of the stylish shops and boutiques housed inside elegant buildings along with a number of cafes for shoppers to rest in. Data was collected under similar weather conditions (average 20$^{\circ}$ degrees), at around 11am.During the data collection process 550,432 data lines of locations traces were collected as well as 5,345 self-report responses.

\section{Discussion}
We explore the use of Voronoi diagrams to map the impact physiological and environmental sensor data have within an urban environment. Voronoi is a computational geometry algorithm that allows the visualisation of data \cite{Dobrin2005}. It works by defining a set of regions called cells. Once the diagram has been created the size of the overall cells give an indication of the density of the area an object is in or the size of the object itself.

\begin{figure}[h]
\centering
  \includegraphics[width= 5.5cm]{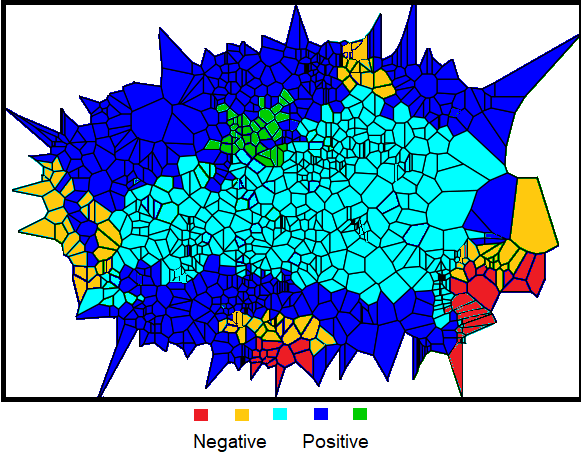}
  \caption{Voronoi showing self-report emotions data within a fragrance and Beauty Shop in Nottingham.}
  \label{fig:bluetored}
\end{figure}

Figure \ref{fig:bluetored} demonstrates self-reported emotions data using the EnvBodySens app within a city environment. To represent the metric value of each record, we apply a linear mapping from the value to the color hue, which forms a blue to red rainbow palette. In particular, the red polygons highlight there was negative sentiment reported by the user. The green polygons aim to highlight where there was a positive sentiment. 

Within the diagram, the negative sentiments, depicted in red polygons, demonstrates where participants' needed to respond to an assistant within a shop. This correlation demonstrates the negative impact on an individual's well-being. We see these results being of particular importance to city planners, to help them design and understand what individual's needs in city centres to help reduce stress and improve individual well-being. Our visualisation tools have enabled the mapping of all collected variables (physiological and environmental) as separate layers.  

\begin{figure}[h]
\centering
  \includegraphics[width= 6cm]{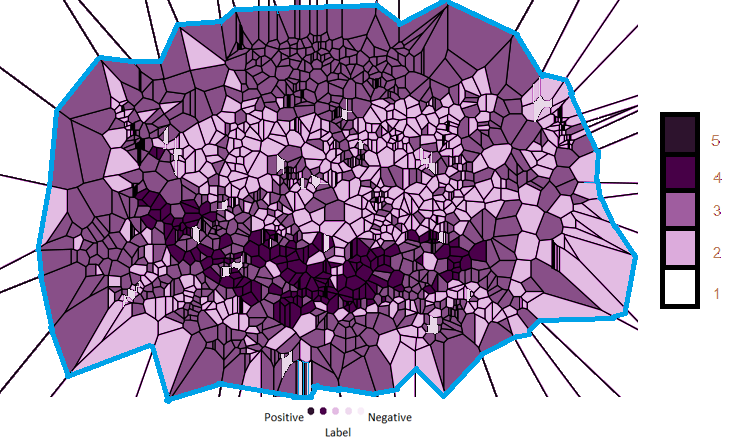}
   \includegraphics[width= 6cm]{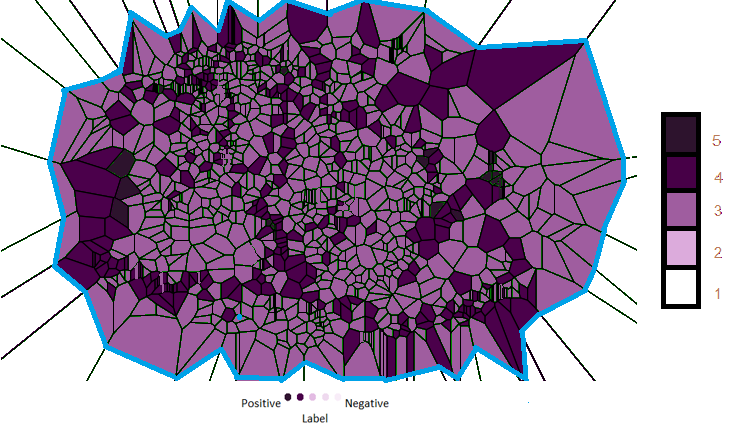}
  \caption{Voronoi presenting the heart rate of participants whilst visiting a popular clothes shop. in Nottingham.}
  \label{fig:purple}
\end{figure}

Figure \ref{fig:purple} shows two Voronoi diagrams displaying the heart rate (HR) when participants entered a range of different urban environments. The darker colours used indicate higher levels of heat rate. Through analysis of sensor data we have discovered a trend when shopping, in that heart rate frequently increases when encountering discounted items in the shop. 

\begin{figure}[h]
\centering
  \includegraphics[height= 7.5cm]{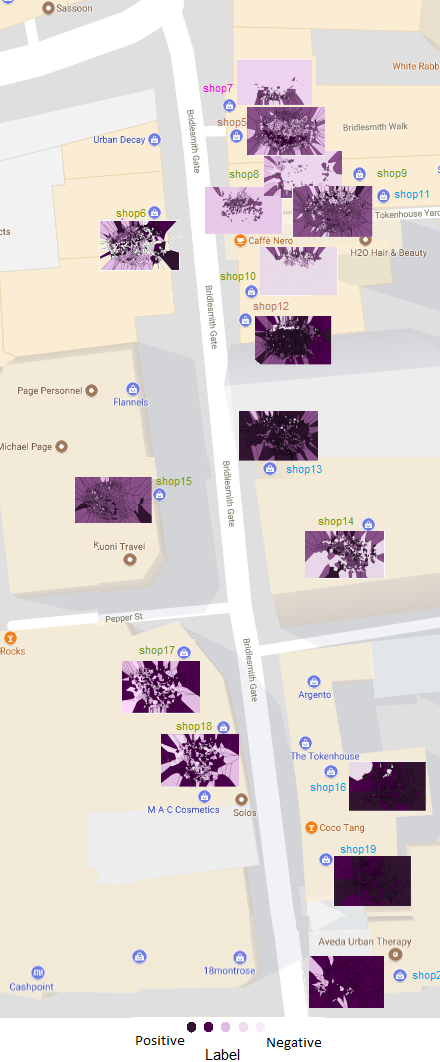}
  \caption{Heart rate Voronoi diagram overlays of different shops in popular Nottingham shopping street.}
  \label{fig:vmap}
\end{figure}

Using Voronoi diagrams to express data visually offers many opportunities to plot correlations and observe patterns. Figure \ref{fig:vmap} presents a map of a street in Nottingham City Centre where Voronoi diagrams have been created using the heart rate and location sensor data. By using location tracking and the utilisation of Voronoi diagrams plotted on to maps has clearly shown the areas where there is an increase in heart rate and stress. 

We have found that these Voronoi charts can be utilised by city planners to explore areas that induce the most stress such as a busy shopping environment and create calming spaces to develop a restorative experience. By enabling people to reduce stress and improve their well-being will have positive impacts on cities urban environments by creating a positive experience, potentially helping to increase the number of visitors and improve shopping experiences.
\section{Conclusion}
We have shown how Voronoi diagrams demonstrate great potential to spatially visualise and explore emotional and physiological sensor data. This spatial exploration enables city planners to develop urban environments that can help to reduce stress and improve well-being. It is hoped that city centres could utilise this data to add calming areas such as green spaces in environments which are considered as high stress. By understanding the environments which induce stress and those that aid relaxation we hope future urban developments can use this method of data visualisation to create restorative experiences in high stress environments to improve mental well-being. In the future, we envision the use of Voronoi charts should continue to be explored in urban environments. With more participants, additional sensor data such as additional environmental measurements (gases, particulate matter, noise) and in more urban city environments will help to evaluate the full potential this form of spatial exploration enables and further understand how urban environments impact mental well-being.

\bibliographystyle{ACM-Reference-Format}
\bibliography{sample-base}

\end{document}